\documentclass[11pt]{article} 

\usepackage{amsmath}
\usepackage{graphicx}

\begin{document}

%---------------------------------------------------------------
% Commands 
%---------------------------------------------------------------

\newcommand{\ket}[1]{| #1 \rangle}
\newcommand{\bra}[1]{\langle #1 |}
\newcommand{\braket}[2]{\langle #1 | #2 \rangle}

\newcommand{\comment}[1]{}

\newcommand{\proof}{\noindent {\bf Proof: }}
\newcommand{\qed}{\hfill{\rule{2mm}{2mm}}}

\newtheorem{Theorem}{Theorem}
\newtheorem{Lemma}{Lemma}
\newtheorem{Hypothesis}{Hypothesis}
\newtheorem{Claim}{Claim}
\newtheorem{Proposition}{Proposition}

\def\up{\Uparrow}
\def\down{\Downarrow}
\def\lt{\Leftarrow}
\def\rt{\Rightarrow}

%---------------------------------------------------------------
% Title
%---------------------------------------------------------------

\title{Search by quantum walks on two-dimensional grid without amplitude amplification
\thanks{This research has been supported by the European Social Fund within the projects ``Support for Doctoral Studies at University of Latvia", 
1DP/1.1.1.2.0/09/APIA/VIAA/044,
FP7 Marie Curie International Reintegration
Grant PIRG02-GA-2007-224886 and FP7 FET-Open project QCS.}}
\author{Andris Ambainis, Art\={u}rs Ba\v{c}kurs, Nikolajs Nahimovs,\\ Raitis Ozols, Alexander Rivosh \\ \\
Faculty of Computing, University of Latvia, \\ Rai\c{n}a bulv. 19, R\={\i}ga, LV-1586, Latvia}
\date{}

\maketitle

%---------------------------------------------------------------
% Abstract
%---------------------------------------------------------------

\textbf{Abstract}.

We study search by quantum walk on a finite two dimensional grid. The algorithm of Ambainis, Kempe, Rivosh \cite{AKR05} $O(\sqrt{N \log{N}})$ steps and finds a marked location with probability $O(1 / \log{N})$ for grid of size $\sqrt{N} \times \sqrt{N}$. This probability is small, thus \cite{AKR05} needs amplitude amplification to get $\Theta(1)$ probability. The amplitude amplification adds an additional $O(\sqrt{\log{N}})$ factor to the number of steps, making it $O(\sqrt{N} \log{N})$.

In this paper, we show that despite a small probability to find a marked location, the probability to be within $O(\sqrt{N})$ neighbourhood (at $O(\sqrt[4]{N})$ distance) of the marked location is $\Theta(1)$. This allows to skip amplitude amplification step and leads to $O(\sqrt{\log{N}})$ speed-up.

We describe the results of numerical experiments, supporting this idea, and we prove this fact analytically.

%---------------------------------------------------------------
% Section: Introduction
%---------------------------------------------------------------

\section{Introduction}

Quantum walks are quantum counterparts of random walks \cite{Amb03, Kem03}. 
They have been useful to design quantum algorithms for a variety of problems \cite{CC+03, Amb04, Sze04, AKR05, MSS05, BS06}. In many of those applications, quantum walks are used as a tool for search.

To solve a search problem using quantum walks, we introduce marked locations corresponding to elements of the search space we want to find. We then perform a quantum walk on search space with one transition rule at unmarked locations and another transition rule at marked locations. If this process is set up properly, it leads to a quantum state in which marked locations have higher probability than unmarked ones. This method of search using quantum walks was first introduced in \cite{SKW03} and has been used many times since then.

In this paper we study the quantum walks on a finite two-dimensional grid according to \cite{AKR05}. It has been shown that after $O(\sqrt{N \log{N}})$ steps a quantum walk on 2D grid with one or two marked locations reaches a state that is significantly different from the state of a quantum walk with no marked location. If this state is measured the probability to obtain a marked location is $O(1/\log{N})$. This probability is small, thus \cite{AKR05} uses amplitude amplification. Amplitude amplification adds an additional $O(\sqrt{\log{N}})$ factor to the number of steps, making it $O(\sqrt{N} \log{N})$.

In case of two-dimensional grid it is logical to examine not only the marked location but also its close neighbourhood. We show that despite small probability to find marked location, the probability to be within $O(\sqrt{N})$ neighbourhood, i.e. at $O(\sqrt[4]{N})$ distance from the marked location, is $\Theta(1)$. This allows us to skip amplitude amplification step and leads to $O(\sqrt{\log{N}})$ speed-up.

The same speed-up has been already achieved by other research groups. Their approaches to this problem are based on modification of the original algorithm \cite{Tul08} or both the algorithm and the structure of the grid \cite{KM+10}.

Our result shows that the improvement of the running time to $O(\sqrt{N} \log{N})$
can be achieved without any modifications to the quantum algorithm, with just a simple
classical post-processing. 
\comment{
We propose a different approach. We keep \cite{AKR05} quantum walk algorithm unchanged but interpret the result of measurement in a different way. This allows us to replace amplitude amplification with classical post processing, which does not increase time complexity of the algorithm. We believe that the proposed approach can be interesting to the reader, especially in the methodological aspect. We also believe that this approach can be used in other quantum walk algorithms.

We provide the reader with results of numerical experiments supporting our idea, as well as present an analytical proof of this fact.
}

%---------------------------------------------------------------
% Section: Quantum walks in two dimensions
%---------------------------------------------------------------

\section{Quantum walks in two dimensions}

Suppose we have $N$ items arranged on a two dimensional lattice of size $\sqrt{N} \times \sqrt{N}$. We will also denote $n=\sqrt{N}$.
The locations on the lattice are labelled by their $x$ and $y$ coordinate as $(x,y)$ for $x,y \in \{ 0, \dots, n-1\}$. We assume that the grid has periodic boundary conditions. For example, going right from a location $(n-1, y)$ on the right edge of the grid leads to the location $(0, y)$ on the left edge of the grid.

\comment{
To introduce quantum version of random walk, it is logical to define basis states $\ket{i, j}$, $i, j \in \{0, \dots ,n-1\}$, and let the state of the quantum walk $\ket{\psi(t)}$ be

\begin{equation}
\label{E1}
\ket{\psi(t)} = \sum_{i,j}{\alpha_i\ket{i,j}}
\end{equation}

\noindent
To evolve the next step of quantum walk $\ket{\psi(t+1)} = U\ket{\psi(t)}$, we use some unitary operator $U$. It is natural to require that $U$ maps each $\ket{i,j}$ to a superposition of $\ket{i,j}$ and the adjacent locations, in a way that is independent of $i$ and $j$. Unfortunately, there is no non-trivial unitary transformations $U$ of this form \cite{AKR05}.
}

To define a quantum walk, we add an additional "coin" register with four states, one for each direction: $\ket{\up}$, $\ket{\down}$, $\ket{\lt}$ and $\ket{\rt}$. At each step we perform a unitary transformation on the extra register and then evolve the system according to the state of the coin register. Thus, the basis states of quantum walk are $\ket{i,j,d}$ for $i,j \in \{0,\dots,n-1\}$, $d \in \{\up, \down, \lt, \rt\}$ and the state of quantum walk is given by:

\begin{align}
\label{E2}
\ket{\psi(t)} = \sum_{i,j} (
& \alpha_{i,j,\up}\ket{i,j,\up} + \alpha_{i,j,\down}\ket{i,j,\down} + \\
\nonumber
& \alpha_{i,j,\lt}\ket{i,j,\lt} + \alpha_{i,j,\rt}\ket{i,j,\rt} )
\end{align}

A step of the coined quantum walk is performed by first applying $I \times C$, where $C$ is unitary transform on the coin register. The most often used transformation on the coin register is the Grover's diffusion transformation $D$:

\begin{equation}
\label{E3}
D = \frac{1}{2} \left( 
\begin{array}{cccc}
-1 & 1 & 1 & 1 \\
1 & -1 & 1 & 1 \\
1 & 1 & -1 & 1 \\
1 & 1 & 1 & -1 
\end{array} \right)
\end{equation}
\\
Then, we apply the shift transformation $S$:
\\
\begin{equation}
\label{E4}
\begin{array}{lcl}
\ket{i,j,\up} & \rightarrow & \ket{i,j-1,\down} \\
\ket{i,j,\down} & \rightarrow & \ket{i,j+1,\up} \\
\ket{i,j,\lt} & \rightarrow & \ket{i-1,j,\rt} \\
\ket{i,j,\rt} & \rightarrow & \ket{i+1,j,\lt}
\end{array}
\end{equation}

Notice that after moving to an adjacent location we change the value of the direction register to the opposite. This is necessary for the quantum walk algorithm of \cite{AKR05} to work.

We start quantum walk in the state 
\[
\ket{\psi(0)} = \frac{1}{2 \sqrt{N}} \sum_{i,j} \big( \ket{i,j,\up} + \ket{i,j,\down} + \ket{i,j,\lt} + \ket{i,j,\rt} \big)
\]

It can be easily verified that the state of the walk stays unchanged, regardless of the number of steps. To use quantum walk as a tool for search, we "mark" some locations. In unmarked locations, we apply the same transformations as above. In marked locations, we apply $-I$ instead of $D$ as the coin flip transformation. The shift transformation remains the same in both marked and unmarked locations.

If there are marked locations, the state of this process starts to deviate from $\ket{\psi(0)}$. It has been shown \cite{AKR05} that after $O(\sqrt{N\log{N}})$ steps the inner product $\braket{\psi(t)}{\psi(0)}$ becomes close to $0$. 

In case of one or two marked locations \cite{AKR05} algorithm finds a marked location with $O(1 / \log{N})$ probability. For multiple marked locations this is not always the case. There exist marked location configurations for which quantum walk fails to find any of marked locations \cite{AR08}.

%---------------------------------------------------------------
% Section: Results
%---------------------------------------------------------------

\section{Results}

In this paper we examine a single marked location case only. However, 
we note that numerical experiments give very similar results 
in the case of multiple marked locations.

Suppose we have an $\sqrt{N} \times \sqrt{N}$ grid with one marked location.
The \cite{AKR05} algorithm takes $O(\sqrt{N \log{N}})$ steps and finds the marked location with $O(1 / \log{N})$ probability. The algorithm then uses amplitude amplification to get $\Theta(1)$ probability. The amplitude amplification adds an additional $O(\sqrt{\log{N}})$ factor to the number of steps, making it $O(\sqrt{N} \log{N})$.

Performing numerical experiments with \cite{AKR05} algorithm, we have noticed that probability to be close to the marked location is much higher than probability to be far from the marked location.
\begin{figure}[h]
\centering
\includegraphics[scale=0.8]{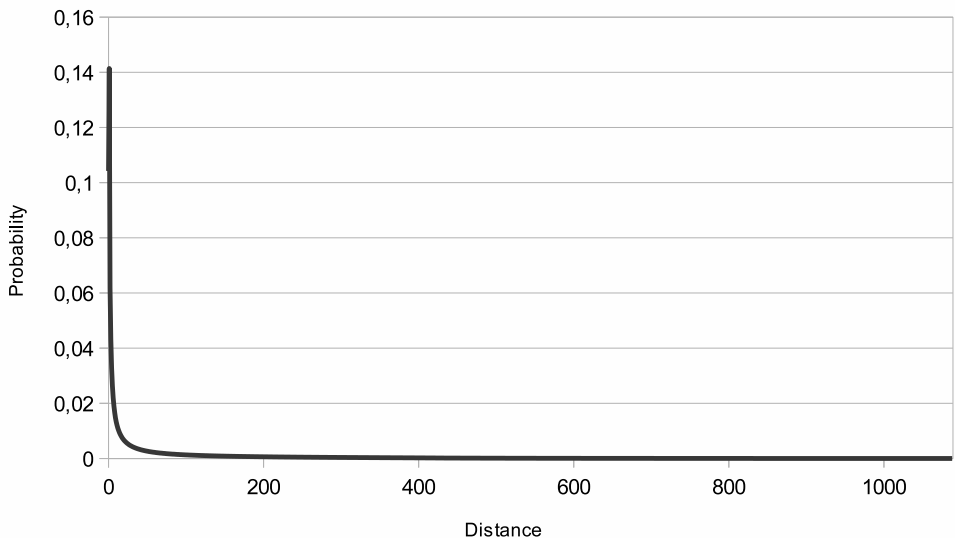}
\caption{Probability by distance, one marked location, grid size $1024 \times 1024$, normal scale.}
\label{fig:Probability_by_distance_normal_scale}
\end{figure}
Figure \ref{fig:Probability_by_distance_normal_scale} shows probability distribution by distance from the marked location for $1024 \times 1024$ grid. 
\begin{figure}[h]
\centering
\includegraphics[scale=0.8]{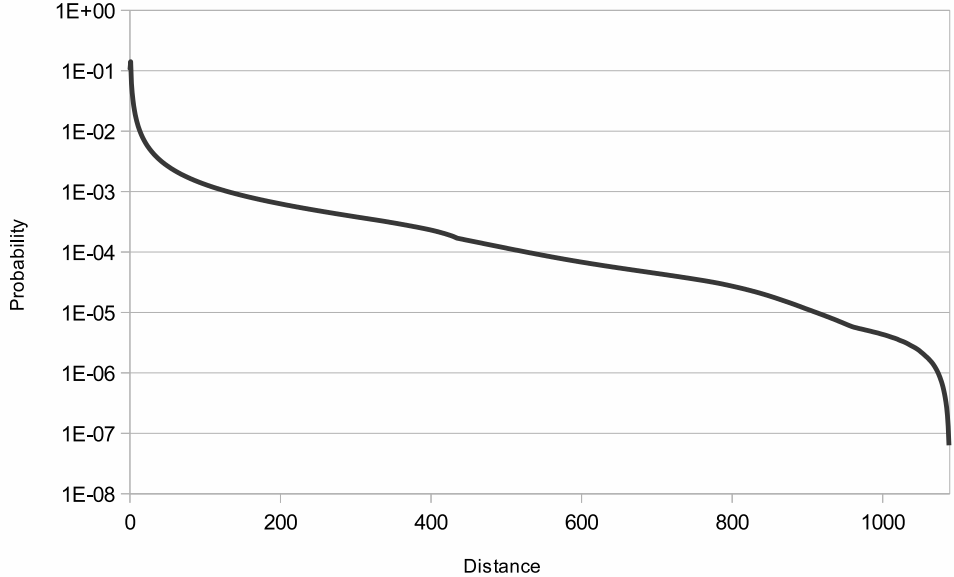}
\caption{Probability by distance, one marked location, grid size $1024 \times 1024$, logarithmic scale.}
\label{fig:Probability_by_distance_log_scale}
\end{figure}
Figure \ref{fig:Probability_by_distance_log_scale} shows the same probability distribution on logarithmic scale.

\begin{figure}[h]
\centering
\includegraphics[scale=0.8]{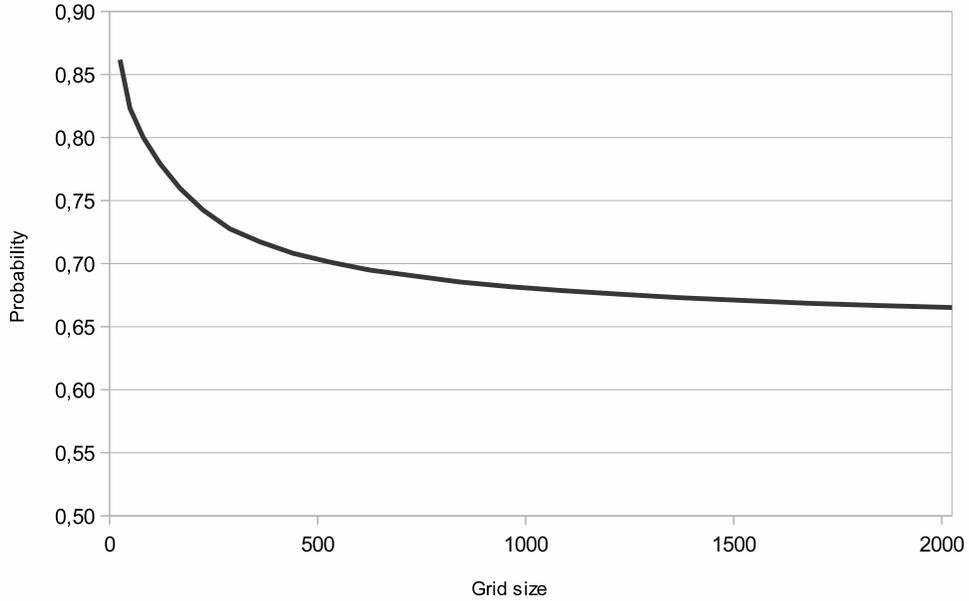}
\caption{Probability to be within $\sqrt{N}$ neibourghood from the marked location.}
\label{fig:Neibourghood_probability}
\end{figure}

We have measured the probability within $O(\sqrt{N})$ neibourghood of the marked location (at $O(\sqrt[4]{N})$ distance) for different grid sizes (figure \ref{fig:Neibourghood_probability}) 
and have made the following conjecture:
\footnote {Another logical choice of the size of the neighbourhood would be $O(\sqrt{N \log{N}})$ - the number of steps of \cite{AKR05} algorithm. 
}
\begin{Hypothesis}
\label{thm:main}
The probability to be within $O(\sqrt{N})$ neighbourhood, i.e. at $O(\sqrt[4]{N})$ distance, of the marked location is $\Theta(1)$. 
\end{Hypothesis}
In the next section we present a strict analytical proof of the conjecture, which in the further discussion will be referred as theorem \ref{thm:main}. 

The theorem allows us to replace amplitude amplification with a classical post processing step. After the measurement we classically check $O(\sqrt{N})$ neighbourhood of the outcome. This requires extra $O(\sqrt{N})$ steps but removes $O(\sqrt{\log{N}})$ factor. Therefore, the running time of the algorithm stays $O(\sqrt{N \log{N}})$.

Before going into details of the proof, we would like to give the reader some  understanding of the final state of the algorithm (state  before the measurement). Denote $Pr[0]$ the probability to find a marked location and $Pr[R]$ the probability to be at distance $R$ from the marked location. For small $R$ values ($R \ll \sqrt{N}$), the numerical experiments indicate that:

$$ Pr[R] \approx \frac{Pr[0]}{R^2} $$

\noindent
There are $4R$ points at the distance $R$ from the marked location (we use Manhattan or $L_1$ distance). Thus, the total probability to be within $\sqrt{N}$ neighbourhood of the marked location is:

$$ 
S = \sum_{R=1}^{\sqrt[4]{N}} 4R \times O \left( \frac{Pr[0]}{R^2} \right) = 
Pr[0] \times\sum_{R=1}^{\sqrt[4]{N}} O \left( \frac{1}{R} \right) = 
Pr[0] \times O(\log{N}) .
$$

\noindent
As probability to find the marked location is $O(1 / \log{N})$, we have 

$$ S = O\left(\frac{1}{\log{N}}\right) \times O(\log{N}) = const .$$

%---------------------------------------------------------------
% Section: Proofs
%---------------------------------------------------------------

\section{Proofs}

In this section, we show

\begin{Theorem}
We can choose $t=O(\sqrt{N \log N})$ so that, if we run a quantum walk with
one marked location $(i, j)$ for $t$ steps and measure the final state,
the probability of obtaining a location $(i', j')$ with $|i-i'|\leq N^{\epsilon}$
and $|j-j'|\leq N^{\epsilon}$ as the measurement result is $\Omega(\epsilon)$
\footnote{Here, $|i-i'|\leq N^{\epsilon}$ and $|j-j'|\leq N^{\epsilon}$ should be interpreted ``modulo $N$": $|i-i'|\leq N^{\epsilon}$
if $(i-i') \bmod N \in\{ -N^{\epsilon}, -N^{\epsilon}+1, \ldots, N^{\epsilon}\}$.}.
\end{Theorem}

The proof of Theorem \ref{thm:main} consists of two steps. First, in Lemma \ref{lem:1},
we derive an approximation
for the state of quantum walk, at the time $t=O(\sqrt{N \log N})$
when the state of quantum walk has the biggest difference from
the starting state. Then, in section \ref{sec:end}, we use this 
approximation to derive our main result, via a sequence of algebraic
transformations and approximations.

\subsection{Approximation of the state of the quantum walk}

Let \[ \ket{\psi}=\sum_{j=0}^{\sqrt{N}-1} \sum_{j'=0}^{\sqrt{N}-1} \sum_{d} 
\alpha^t_{j, j', d} \ket{j, j',d} \] be the state of the quantum walk
after $t$ steps.

\begin{Lemma}
\label{lem:1}
We can choose $t=O(\sqrt{N \log N})$ so that for any set 
\[ S\subseteq \{0, ..., \sqrt{N}-1\}^2, \]
we have
\[ \sum_{(j, j') \in S} |\alpha^t_{j, j', \up}|^2   
\geq C^2\sum_{(j, j')\in S} (f(j, j')-f(j-1, j'))^2 +o(1) \]
where 
\[ f(j, j')= 
\sum_{(k, l)\neq (0, 0)} \frac{1}{2- \cos \frac{2 k \pi}{\sqrt{N}} - \cos \frac{2l \pi}{\sqrt{N}}} 
w^{kj + lj'}, 
\]
$w=e^{2\pi i \over \sqrt N}$ and $C=\Theta(\frac{1}{N \sqrt{\log N}})$.
\end{Lemma}

\proof
We will repeatedly use the following lemma.
\begin{Lemma}
\cite{BV}
\label{lem:bv}
Let $\ket{\psi} = \sum_{i=1}^m \alpha_i \ket{i}$ and
$\ket{\psi'} = \sum_{i=1}^m \beta_i \ket{i}$.
Then, for any set $S\subseteq \{1, 2, \ldots, m\}$,
\[ \sum_{i\in S} \left| |\alpha_i|^2 - |\beta_i |^2 \right| 
\leq 2 \|\psi-\psi'\| .\]
\end{Lemma}

We recast the algorithm for search on the grid as 
an instance of an {\em abstract search algorithm} \cite{AKR05}.
An abstract search algorithm consists of two unitary
transformations $U_1$ and $U_2$ and two states 
$\ket{\psi_{start}}$ and $\ket{\psi_{good}}$.
We require the following properties:
\begin{enumerate}
\item
$U_1=I-2\ket{\psi_{good}}\bra{\psi_{good}}$ (in other words, $U_1\ket{\psi_{good}}=-
\ket{\psi_{good}}$ and, if $\ket{\psi}$ is orthogonal to $\ket{\psi_{good}}$,
then $U_1\ket{\psi}=\ket{\psi}$);
\item
$U_2\ket{\psi_{start}}=\ket{\psi_{start}}$ for some state $\ket{\psi_{start}}$ with 
real amplitudes and there is no other eigenvector with eigenvalue 1;
\item
$U_2$ is described by a real unitary matrix.
\end{enumerate}
The abstract search algorithm applies the unitary transformation
$(U_2 U_1)^T$ to the starting state $\ket{\psi_{start}}$.
We claim that under certain constraints its final state
$(U_2 U_1)^T\ket{\psi_{start}}$ has a sufficiently large inner product with
$\ket{\psi_{good}}$.

For the quantum walk on $\sqrt{N}\times \sqrt{N}$ grid, 
\[ \ket{\psi_{good}} = \frac{1}{2} \ket{i, j, \up} + \frac{1}{2} \ket{i, j, \down}
+ \frac{1}{2} \ket{i, j, \lt} + \frac{1}{2} \ket{i, j, \rt} ,\]
where $i, j$ is the marked location and
\[ \ket{\psi_{start}} = \frac{1}{2\sqrt{N}} \sum_{i, j=0}^{\sqrt N-1} \left( \ket{i, j, \up} + \ket{i, j, \down} + \ket{i, j, \lt} + \ket{i, j, \rt} \right).\]

Since $U_2$ is described by a real-value matrix, its eigenvectors (with eigenvalues that are not 
1 or -1) can be divided into pairs: 
$\ket{\Phi_j^+}$ and $\ket{\Phi_j^-}$, with
eigenvalues $e^{i\theta_j}$ and $e^{-i\theta_j}$, respectively.
In the case of the walk on the 2-dimensional grid, these eigenvalues were calculated in Claim 6 of \cite{AKR05}: 

\begin{Claim}
Quantum walk on the 2-dimensional grid with no marked locations has
$N-1$ pairs of eigenvalues $e^{-i\theta_j}$ that are not equal to 1 or -1.
These values can be indexed by pairs $(k, l)$, $k, l\in\{0, 1, \ldots, \sqrt{N}-1\}$, $(k, l)\neq (0, 0)$.
The corresponding eigenvalues are equal to $e^{\pm i\theta_{k, l}}$, where $\theta_{k, l}$ satisfies
$\cos \theta_{k, l} = \frac{1}{2} ( \cos \frac{2 \pi k}{\sqrt{N}} + \cos \frac{2 \pi l}{\sqrt{N}} )$.
\end{Claim}

We use $\ket{\Phi_{k, l}^+}$ and $\ket{\Phi_{k, l}^-}$ to denote the corresponding eigenvectors.
According to \cite[pages 3-4]{MPA10},
these eigenvectors are equal to $\ket{\Phi^+_{k, l}}=\ket{\xi_k}\otimes \ket{\xi_l}\otimes \ket{v^+_{k, l}}$,
$\ket{\Phi^-_{k, l}}=\ket{\xi_k}\otimes \ket{\xi_l}\otimes \ket{v^-_{k, l}}$ 
where $\ket{\xi_k}=\sum_{i=0}^{\sqrt{N}-1} w^{k i} \frac{1}{\sqrt[4]{N}} \ket{i}$,
\[ \ket{v^+_{k, l}} = \frac{i}{2\sqrt{2} \sin \theta_{k, l}} \left[ \begin{array}{c} e^{-i \theta_{k, l}} - w^k \\  e^{-i \theta_{k, l}} - w^{-k} \\
 e^{-i \theta_{k, l}} - w^l \\  e^{-i \theta_{k, l}} - w^{-l} \end{array} \right] , \mbox{~~~}
\ket{v^-_{k, l}} = \frac{i}{2\sqrt{2} \sin \theta_{k, l}} \left[ \begin{array}{c} w^k - e^{i \theta_{k, l}}  \\ w^{-k} - e^{i \theta_{k, l}}  \\
w^l - e^{i \theta_{k, l}}  \\ w^{-l} - e^{i \theta_{k, l}}  \end{array} \right] .\]
The order of directions for the coin register is: $\ket{\down}$, $\ket{\up}$, $\ket{\rt}$, $\ket{\lt}$.
The sign of $\ket{v^{-}_{k, l}}$ has been adjusted so that
\begin{equation}
\label{eq:12} \frac{1}{\sqrt{2}} \ket{\Phi^+_{k, l}} + \frac{1}{\sqrt{2}} \ket{\Phi^-_{k, l}} = \ket{\xi_k}\otimes \ket{\xi_l}\otimes\ket{\psi_0} 
\end{equation}
where $\ket{\psi_0}={1\over 2}\ket{\down}+{1\over 2}\ket{\up}+{1\over 2}\ket{\rt}+{1\over 2}\ket{\lt}.$

We can assume that $\ket{\psi_{good}}=\ket{0}\otimes\ket{0}\otimes\ket{\psi_0}$. This gives us an expression of $\ket{\psi_{good}}$ in terms of the eigenvectors of $U_2$:
\[
\ket{\psi_{good}}=\frac{1}{\sqrt{N}} \sum_{k, l} \ket{\xi_k} \otimes \ket{\xi_l}\otimes\ket{\psi_0} \] \[ =
{1\over \sqrt N}\ket{\psi_{start}}+\sum_{(k, l)\neq (0, 0)}
\left( \frac{1}{\sqrt{2N}} \ket{\Phi^+_{k, l}} + \frac{1}{\sqrt{2N}} \ket{\Phi^-_{k, l}} \right) .
\]

Using the results from \cite{AKR05}, we can transform this into an expression for the final state of our quatum search algorithm.
According to the first big equation in the proof of Lemma 5 in \cite{AKR05},
after $t=O(\sqrt{N \log N})$ steps,
we get a final state $\ket{\psi}$ such that 
$\|\ket{\psi} - \ket{\phi_{final}}\|=o(1)$,
where $\ket{\phi_{final}}=\frac{\ket{\phi'_{final}}}{\|\phi'_{final}\|}$ and
\begin{equation}
\label{eq:final0}
\ket{\phi'_{final}} = \frac{2}{\sqrt{N}} \ket{\psi_{start}}+
{1\over \sqrt{2N}}\sum_{(k, l)\neq (0, 0)}
a_{k, l} \ket{\Phi^+_{k, l}} +  b_{k, l} \ket{\Phi^-_{k, l}} 
\end{equation}
and
\[ a_{k, l} = 1 + \frac{i}{2} \cot \frac{\alpha+\theta_{k, l}}{2} + \frac{i}{2} \cot \frac{-\alpha+\theta_{k, l}}{2} ,\]
\[ b_{k, l} = 1 + \frac{i}{2} \cot \frac{\alpha-\theta_{k, l}}{2} + \frac{i}{2} \cot \frac{-\alpha-\theta_{k, l}}{2} .\]
We now replace $\sum_{(j, j') \in S} |\alpha^t_{j, j', d}|^2$
by the corresponding sum of squares of amplitudes for the state $\ket{\phi_{final}}$.
By Lemma \ref{lem:bv}, this changes the sum by an amount that is $o(1)$.

From \cite{AKR05}, we have $\alpha=\Theta(\frac{1}{\sqrt{N \log N}})$,
$\min \theta_{k, l} = \Theta(\frac{1}{\sqrt{N}})$ and
$\max \theta_{k, l} = \frac{\pi} - \Theta(\frac{1}{\sqrt{N}})$.
Hence, we have  
$\pm \alpha+\theta_{k, l}= (1+o(1)) \theta_{k, l}$ 
and we get 
\[ \ket{\phi'_{final}} = \frac{1}{\sqrt{N}} \ket{\psi_{start}}+\sum_{(k, l)\neq (0, 0)}
\frac{1}{\sqrt{2N}} \left( 1 + i (1+o(1)) \cot \frac{\theta_{k, l}}{2} \right) \ket{\Phi^+_{k, l}} +  \]
\begin{equation}
\label{eq:final1}
\frac{1}{\sqrt{2N}} \left( 1 - i (1+o(1)) \cot \frac{\theta_{k, l}}{2} \right)\ket{\Phi^-_{k, l}} .
\end{equation}
This means that $\|\ket{\psi_{final}}-\ket{\phi_{final}}\| = o(1)$ where
$\ket{\psi_{final}}=\frac{\ket{\psi'_{final}}}{\|\psi'_{final}\|}$ and
\begin{equation}
\label{eq:final2}
\ket{\psi'_{final}} = \ket{\psi_{good}} + \sum_{(k, l)\neq (0, 0)}
\frac{1}{\sqrt{2N}}  i \cot \frac{\theta_{k, l}}{2}  
\left( \ket{\Phi^+_{k, l}} -  \ket{\Phi^-_{k, l}} \right) .
\end{equation}
Again, we can replace a sum of squares of amplitudes for the state $\ket{\phi_{final}}$ 
by the corresponding sum for $\ket{\psi_{final}}$ and,
by Lemma \ref{lem:bv}, the sum changes by an amount that is $o(1)$.

We now estimate the amplitude of $\ket{j, j', \up}$ in $\ket{\psi_{final}}$.
We assume that $(j, j') \neq (0, 0)$.
Then, the amplitude of $\ket{j, j', \up}$ in $\ket{\psi_{good}}$ is 0.
Hence, we can evaluate the amplitude of $\ket{j, j', \up}$ in
\begin{equation}
\label{eq:apx} 
\sum_{(k, l)\neq (0, 0)}
\frac{1}{\sqrt{2N}}  i \cot \frac{\theta_{k, l}}{2}  
( \ket{\Phi^+_{k, l}} -  \ket{\Phi^-_{k, l}} )
\end{equation}
and then divide the result by $\Theta(\sqrt{\log N})$, because 
$\| \psi'_{final} \| = \Theta(\sqrt{\log N})$.

From the definitions of $\ket{\Phi^{\pm}_{k, l}}$ and $\ket{v^{\pm}_{k, l}}$,
\[ \frac{1}{\sqrt{2}} \ket{v^+_{k, l}} - \frac{1}{\sqrt{2}} \ket{v^-_{k, l}} =
\frac{i}{4 \sin \theta_{k, l}}  \left[ \begin{array}{c} 2 \cos \theta_{k, l} - 2 w^k  \\  2 \cos \theta_{k, l} - 2 w^{-k}  \\
 2 \cos \theta_{k, l} - 2 w^l \\  2 \cos \theta_{k, l} - 2 w^{-l} \end{array}  \right] \]
Therefore, the amplitude of $\ket{\up}$ in this state is $\frac{i}{2\sin \theta_{k, l}} (\cos \theta_{k, l} - w^{-k})$.
The amplitude of $\ket{j}$ in $\ket{\xi_k}$ is $\frac{1}{\sqrt[4]{N}} w^{kj}$. 
The amplitude of $\ket{j'}$ in $\ket{\xi_l}$ is $\frac{1}{\sqrt[4]{N}} w^{lj'}$. 
Therefore, the amplitude of $\ket{j, j', \up}$ in 
$\frac{1}{\sqrt{2}} \ket{\Phi^+_{k, l}} - \frac{1}{\sqrt{2}} \ket{\Phi^-_{k, l}}$ is
\[ \frac{1}{\sqrt{N}} w^{kj+lj'} \frac{i}{2\sin \theta_{k, l}} (\cos \theta_{k, l} - w^{-k}) \]
and the amplitude of $\ket{j, j', \up}$ in (\ref{eq:apx}) is 
\[ \frac{1}{\sqrt{2} N}
\sum_{(k, l)\neq (0, 0)}
 i \cot \frac{\theta_j}{2}  \cdot \frac{i}{2 \sin \theta_{k, l}} (\cos \theta_{k, l} - w^{-k}) w^{kj+lj'}.\]
By using 
$\sin \theta_{k, l}=2\sin \frac{\theta_{k, l}}{2} \cos \frac{\theta_{k, l}}{2}$, 
we get that the amplitude of $\ket{j, j', \up}$ is
\[ 
{1\over \sqrt 2}\sum_{(k, l)\neq (0, 0)} \frac{1}{4 N} \left( 
- \frac{\cos \theta_{k, l}}{\sin^2 \frac{\theta_{k, l}}{2} } w^{kj + lj'}  +
\frac{1}{\sin^2 \frac{\theta_{k, l}}{2}} w^{k(j-1) + lj'}
\right) \]
\begin{equation}
\label{eq:2} 
= 
{1\over \sqrt 2}\sum_{(k, l)\neq (0, 0)} \frac{1}{4 N} \left( 
2 w^{kj + lj'} - \frac{1}{\sin^2 \frac{\theta_{k, l}}{2} } 
(w^{kj + lj'} - w^{k(j-1) + lj'}) \right)
,\end{equation}
with the equality following from $\cos 2x = 1- 2 \sin^2 x$.

We can decompose the sum into two sums, one over all the first components, 
one over all the second components. 
The first component of the sum in (\ref{eq:2}) is close to 0 and, therefore, can be omitted. Hence, we get
that the amplitude of $\ket{j, j', \up}$ in the unnormalized state
$\ket{\psi'_{final}}$ can be approximated by
\[ {1\over \sqrt 2}\sum_{(k, l)\neq (0, 0)} \frac{1}{4N} \frac{1}{\sin^2 \frac{\theta_{k, l}}{2}} 
(- w^{kj + lj'} + w^{k(j-1) + lj'}) = \Theta\left( \frac{1}{N} \right) \cdot (f(j-1, j')-f(j, j')) .
\]
To obtain the amplitude of $\ket{j, j', \up}$ in $\ket{\psi_{final}}$, 
this should be divided by $\|\psi'_{final}\|$ which is of the order $\Theta(\sqrt{\log N})$.
This implies Lemma \ref{lem:1}.
\qed

%---------------------------------------------------------------
% Section: Bounds on the probability of being close to the marked location
%---------------------------------------------------------------

\section{Bounds on the probability of being close to the marked location}
\label{sec:end}

We start by performing some rearrangements in the expression $f(j, j')$.

Let $n=\sqrt N$ and $S$ be the set of all pairs $(k, l)$ such as $k, l \in\{0, 1, \ldots, n-1\}$, except for $(0, 0)$.
We consider
\[ f(j, j') = 
\sum_{(k, l)\in S} \frac{1}{2- \cos \frac{2 k \pi}{n} - \cos \frac{2l \pi}{n}} 
w^{kj + lj'} \] \begin{equation}
\label{eq:1}=
\sum_{(k, l)\in S} \frac{\cos \frac{2(kj+lj')\pi}{n} + \sin \frac{2(kj+lj')\pi}{n} i
}{2- \cos \frac{2 k \pi}{n} - \cos \frac{2l \pi}{n}}
.\end{equation}
Since the cosine function is periodic with period $2\pi$, we have $\cos \frac{2l \pi}{n}=
\cos \frac{2(l-N) \pi}{n}$. Hence, we can replace the summation over $S$ by the summation
over 
\[ S' = \left\{ (k, l) | k, l \in\left\{-\left\lfloor \frac{n}{2} \right\rfloor, 1, \ldots, \left\lfloor \frac{n}{2} -1 \right\rfloor\right\}\right\}\setminus \{(0,0)\} .\]
This implies that the imaginary part of (\ref{eq:1}) cancels out because terms in the sum can be paired up
so that, in each pair, the imaginary part in both terms has the same absolute value but opposite sign. Namely:
\begin{itemize}
\item
If none of $k, l, -k$ and $-l$ is equal to $\frac{n}{2}$, we pair up $(k, l)$ with $(-k, -l)$.
\item
If none of $k$ and $-k$ is equal to 0 or $\frac{n}{2}$, we pair up $(-\frac{n}{2}, k)$ with $(-\frac{n}{2}, -k)$
and $(k, -\frac{n}{2})$ with $(-k, -\frac{n}{2})$.
\item
The terms $(-\frac{n}{2}, 0)$, $(0, -\frac{n}{2})$ and $(-\frac{n}{2}, -\frac{n}{2})$ are left without a pair. 
This does not affect the argument because the imaginary part is equal to 0 in those terms.
\end{itemize}
Hence, we have
\[  
f(j,j')
=
\sum_{(k, l)\in S'} \frac{\cos \frac{2(kj+lj')\pi}{n}}{2- \cos \frac{2 k \pi}{n} - \cos \frac{2l \pi}{n}} .\]

We define a function $g(j,j')=f(j,j')-f(j-1,j')$. 
By Lemma \ref{lem:1}, $C g(j, j')$ is a good approximation
for the amplitude of $\ket{j, j', \up}$ in the state of the quantum
walk after $t=O(\sqrt{N \log N})$ steps.

\begin{Lemma}
\label{lem:AmpSum}
$$
\sum_{0<j',j<M}g^2(j,j')=\Omega(n^2\ln M)
$$
where $M=n^\epsilon$ and $\epsilon=\Omega(1)$, and $\epsilon=1-\Omega(1)$.
\end{Lemma}

Together with Lemma \ref{lem:1}, this implies that the sum of amplitudes 
of $\ket{j, j', \up}$, $0<j',j<M$ is $\Omega(\frac{\log M}{\log n}) - o(1)$.
Since $\frac{\log M}{\log N}=\epsilon$, this would complete the proof of Theorem 1.

\proof [of Lemma \ref{lem:AmpSum}]
We introduce a function
$$
R(M',M'',k)=\sum_{l=M'+1}^{M''}g^2(l,k)
$$
where $M''>M'>k$ and $M''=\alpha M'$ for some $\alpha$.

\begin{Claim}
\label{claim:squares}
$|f(j, j')-\frac{n^2}{2\pi^2}f'(j,j')| = O(n^2)$ where 
\[ f'(j,j') = \sum_{(k, l)\in S'} \frac{\cos \frac{2(kj+lj')\pi}{n}}{k^2+l^2}. \]
\end{Claim}

\begin{Claim}
\label{claim:approx}
Let $j'=j\beta$ where $0<\beta\leq 1$ and $j=n^\epsilon$, and $\epsilon=\Omega(1)$, and $\epsilon=1-\Omega(1)$.

The following equality holds:
$$f'(j,j')={\pi \over 2}\ln{n \over j} + O(1).$$
\end{Claim}

Given these two claims, we now complete the proof of Lemma \ref{lem:AmpSum}.
From the inequality of quadratic and arithmetic means, we get
$$
R(M',M'',k) \geq {\left(f(M'',k)-f(M',k)\right)^2 \over M''-M'}
$$
$$
={\left({n^2 \over 4\pi}\ln{n \over M''}-{n^2 \over 4\pi}\ln{n \over M'} + O(n^2)\right)^2\over M''-M'}
$$
$$
={\left({n^2 \over 4\pi}\ln \alpha + O(n^2)\right)^2 \over (\alpha-1)M'}
$$
$$
={\Omega(n^2)\over M'}
$$
where the first equality follows from $M'', M'>k$ and Claims \ref{claim:squares} and \ref{claim:approx}. The last equality holds if we choose an $\alpha$ large enough that ${n^2 \over 4\pi}\ln \alpha + O(n^2)=\Omega(n^2)$. 

We introduce a notation
$$
P(M')=\sum_{l=0}^{M'-1}R(M',\alpha M',l).
$$

From $R(M',M'',k)={\Omega(n^2)\over M'}$ we get $P(M')=\Omega(n^2)$. We obtain the following lower bound:
$$
\sum_{0<j',j<M}g^2(j,j')>\sum_{0<j'<j<M}g^2(j,j')
$$
$$
>\sum_{l=1}^{\log_{\alpha}\sqrt M}P\left({M\over \alpha^l}\right)=\Omega\left(n^2\log_{\alpha}\sqrt M\right)=\Omega(n^2\ln M).
$$
\qed

\proof [of Claim \ref{claim:squares}] % subsection?

We have
\[ |f(j, j') - \frac{n^2}{2\pi^2}f'(j, j')| \]
\[ \leq \sum_{(k, l)\in S'} \left| \cos \frac{2(kj+lj')\pi}{n} \right|
\cdot 
\left| \frac{1}{2- \cos \frac{2 k \pi}{n} - \cos \frac{2l \pi}{n}} - 
\frac{n^2}{2\pi^2(k^2+l^2)} \right| .\]
The claim now follows from $|S'|=n^2-1$, $|\cos x| \leq 1$ and  
\[ \left| \frac{1}{2- \cos \frac{2 k \pi}{n} - \cos \frac{2l \pi}{n}} - 
\frac{n^2}{2\pi^2(k^2+l^2)} \right| \leq \frac{1}{2} .\]
To prove the last inequality, we first rewrite 
\[ \frac{1}{2- \cos \frac{2 k \pi}{n} - \cos \frac{2l \pi}{n}} = 
\frac{1}{2 (\sin^2 \frac{k \pi}{n} + \sin^2 \frac{l \pi}{n} )} .\]
We have $x - \frac{x^3}{6} \leq \sin x \leq x$ for all $x\in[0, \pi]$.
This implies $x^2 - \frac{x^4}{3} \leq \sin^2x \leq x^2$.
Hence, we have
\[ \left| \frac{1}{2 (\sin^2 \frac{k \pi}{n} + \sin^2 \frac{l \pi}{n} )} - 
\frac{1}{2((\frac{k \pi}{n})^2+(\frac{l \pi}{n})^2)} \right| 
 = 
\frac{(\frac{k \pi}{n})^2+(\frac{l \pi}{n})^2- (\sin^2 \frac{k \pi}{n} + \sin^2 \frac{l \pi}{n})  }{2((\frac{k \pi}{n})^2+(\frac{l \pi}{n})^2)(\sin^2 \frac{k \pi}{n} + \sin^2 \frac{l \pi}{n} )}
\]
\[ \leq
\frac{(\frac{k \pi}{n})^4+(\frac{l \pi}{n})^4}{6 ((\frac{k \pi}{n})^2+(\frac{l \pi}{n})^2)\left((\frac{k \pi}{n})^2+(\frac{l \pi}{n})^2-\frac{(\frac{k \pi}{n})^4+(\frac{l \pi}{n})^4}{3}\right)} 
\leq \frac{1}{2} \]
where the last inequality follows from
$$
\frac{a^2+b^2}{(a+b)\left(a+b-{a^2+b^2\over 3}\right)}\leq 3
$$
which holds for $0\leq a,b \leq ({\pi \over 2})^2$.
\qed

\proof[of Claim \ref{claim:approx}] % subsection?

We will use the notation $\alpha={2\pi \over n}$.

The following equalities hold
$$
\sum_{(k,l) \in  S'}{\cos \alpha(kj+lj') \over k^2+l^2}=\sum_{(k,l) \in  S'}{\cos \alpha j(k+l\beta) \over k^2+l^2}
$$
\begin{equation}
\label{approxEq1}
=\sum_{
\begin{array}{c}
k+l\beta\leq n \\
k,l \in Z^{0+} \\
(k,l) \neq (0,0)
\end{array}
}{\cos \alpha j(k+l\beta) \over k^2+l^2} + O(1).
\end{equation}
The last equality holds because it lacks some summands, with absolute value of their sum bounded above by
$$
\sum_{
\begin{array}{c}
(k,l) \in  S' \\
k+l>n
\end{array}
}
{1 \over k^2+l^2}=O(1).
$$
It also has some new summands, with absolute value of their sum bounded above by
$$
\sum_{
\begin{array}{c}
l>n \\
0<k<n \\
k,l \in Z^{0+} 
\end{array}
}
{1 \over k^2+l^2}=O(1).
$$

We will use the notation $k'=k+\lceil l\beta\rceil-l\beta$. We replace the sum \eqref{approxEq1} (without the asymptotic) with
\begin{equation}
\label{approxEq2}
\sum_{
\begin{array}{c}
k+l\beta\leq n \\
k,l \in Z^{0+} \\
(k,l) \neq (0,0)
\end{array}
}{\cos \alpha j( k'+l\beta) \over k^2+l^2}.
\end{equation}

The error because of the replacement is
$$
2\pi n^{\epsilon-1}
\sum_{
\begin{array}{c}
k+l\beta\leq n \\
k,l \in Z^{0+} \\
(k,l) \neq (0,0)
\end{array}
}{1 \over k^2+l^2}\leq
2\pi n^{\epsilon-1}
\sum_{
\begin{array}{c}
k,l \in Z^{0+} \\
(k,l) \neq (0,0) \\
0 \leq k \leq n \\
l \geq 0
\end{array}
}{1 \over k^2+l^2}
$$
$$
=2\pi n^{\epsilon-1} O(\ln n)=o(1)
$$
where we used the fact that $|\cos \alpha j( k'+l\beta)-\cos \alpha j( k+l\beta)|\leq 2\pi n^{\epsilon-1}$.

We replace the sum \eqref{approxEq2} with
\begin{equation}
\label{approxEq3}
\sum_{
\begin{array}{c}
k+l\beta\leq n \\
k,l \in Z^{0+} \\
(k,l) \neq (0,0)
\end{array}
}{\cos \alpha j( k'+l\beta) \over (k')^2+l^2}.
\end{equation}

The error of the last replacement is
$$
\left|
\sum_{
\begin{array}{c}
k+l\beta\leq n \\
k,l \in Z^{0+} \\
(k,l) \neq (0,0)
\end{array}
}
\left(
{\cos \alpha j(k'+l\beta) \over(k')^2+l^2}-
{\cos \alpha j(k'+l\beta) \over k^2+l^2}
\right)
\right|
$$
$$
\leq\sum_{
\begin{array}{c}
k,l \in Z^{0+} \\
(k,l) \neq (0,0)
\end{array}
}
{ (k')^2-k^2\over (k^2+l^2)^2}\leq
\sum_{
\begin{array}{c}
k,l \in Z^{0+} \\
(k,l) \neq (0,0)
\end{array}
}
{ 2k+1\over (k^2+l^2)^2}
$$
$$
\leq
3\sum_{
\begin{array}{c}
k,l \in Z^{0+} \\
(k,l) \neq (0,0)
\end{array}
}
{ k+l\over (k^2+l^2)^2}\leq
12\sum_{
\begin{array}{c}
k,l \in Z^{0+} \\
(k,l) \neq (0,0)
\end{array}
}
{ 1\over (k+l)^3}=O(1).
$$

We replace the sum \eqref{approxEq3} with
\begin{equation}
\label{approxEq4}
\sum_{
\begin{array}{c}
k+l\beta\leq n \\
k,l \in Z^{0+} \\
(k,l) \neq (0,0)
\end{array}
}\cos \left(\alpha j( k'+l\beta)\right) {1\over \beta}\int_{-{\beta \over 2}}^{\beta \over 2}{dt \over (k'-t)^2+\left(l+{t \over \beta}\right)^2}.
\end{equation}

Bacause of the last replacement the error in a fixed summand  is
$$
\left|
{1\over (k')^2+l^2}- {1\over \beta}\int_{-{\beta \over 2}}^{\beta \over 2}{dt \over (k'-t)^2+\left(l+{t \over \beta}\right)^2}
\right|=
$$
$$
\left|
{1\over (k')^2+l^2}- {\arctan\left( {k'+l\beta \over (k')^2+l^2-{1+\beta^2\over 4}} \right)\over k'+l\beta}
\right|.
$$

By using $x-{x^3 \over 3} < \arctan x < x$ that holds for all $x>0$ we bound the error from above by
$$
\left|
{-{1+\beta^2\over 4}\over ((k')^2+l^2)((k')^2+l^2-{1+\beta^2 \over 4})}
\right|
+
\left|
{\left({k'+l\beta\over (k')^2+l^2-{1+\beta^2 \over 4}}\right)^3 \over 3(k'+l\beta)}
\right|.
$$

By using the inequalities $(k')^2+l^2\geq{1\over 2}(k'+l)^2$ and $(k')^2+l^2-{1\over 2}\geq{1\over 4}(k'+l)^2$ which hold if $k+l\geq 1$ and $k,l\in Z^{0+}$, and $0<\beta\leq 1$, we obtain the following upper bound of the error:
$$
{4\over (k'+l)^4}+{64\over 3(k'+l)^4}={76\over 3(k'+l)^4}\leq{76\over 3(k+l)^4}.
$$

Thus, the error made in \eqref{approxEq4} can be bounded from above by
$$
\sum_{
\begin{array}{c}
k+l\beta\leq n \\
k,l \in Z^{0+} \\
(k,l) \neq (0,0)
\end{array}
}{76\over 3(k+l)^4}=O(1).
$$

We replace \eqref{approxEq4} with
$$
{1\over \beta}\sum_{s=1}^n \cos \alpha j s \int_{0}^{s}{dk \over k^2+\left({s-k\over \beta}\right)^2}.
$$

We grouped summands with equal cosine arguments. We also altered integration limits to obtain an integral on the interval $[0,s]$. The error made in this step can be bounded from above by
$$
\sum_{s=1}^n {1\over s^2}=O(1).
$$

By using $\int_{0}^{s}{dk \over k^2+\left({s-k\over \beta}\right)^2}={\beta \pi \over 2 s}$ we obtain the following sum:
\begin{equation}
\label{approxEq5}
{\pi\over 2}\sum_{s=1}^n {\cos \alpha j s \over s}.
\end{equation}

\begin{Proposition}
\label{prop:sum2}
Let $j=n^\epsilon$ and $\epsilon=\Omega(1)$, and $\epsilon=1-\Omega(1)$.

The following equality holds:
$$
\sum_{k=1}^{n}{\cos\left({2\pi \over n} j k\right) \over k} = (1-\epsilon)\ln n + O(1).
$$
\end{Proposition}

Now Proposition \ref{prop:sum2} gives us that \eqref{approxEq5} is equal to
$$
{\pi \over 2}\ln{n \over j} + O(1).
$$
\qed

\proof [of Proposition \ref{prop:sum2}]

We can rewrite the sum $\sum_{k=1}^{n}{\cos\left({2 \pi \over n} n^\epsilon k\right) \over k}$ in the following way:
$$
\sum_{k=1}^{\lfloor n^{1-\epsilon}\rfloor}{\cos\left(2\pi n^{\epsilon-1} k\right) \over k}+\sum_{l=1}^{\lfloor n^\epsilon \rfloor-1}\sum_{t=\lfloor n^{1-\epsilon} l\rfloor+1}^{\lfloor n^{1-\epsilon}(l+1)\rfloor}{\cos\left(2\pi n^{\epsilon-1}t\right)\over t}
$$
\begin{equation}
\label{sumParts}
+\sum_{k=\lfloor n^{1-\epsilon}\lfloor n^\epsilon\rfloor\rfloor+1}^n {\cos\left(2\pi n^{\epsilon-1}k\right)\over k}
\end{equation}

\begin{Proposition}
\label{prop:sum1}
$$\sum_{k=1}^{n} {\cos\left({2\pi \over n} k\right) \over k} = \ln n + O(1).$$
\end{Proposition}

\proof

The proposition follows from
$$
\sum_{k=1}^{n} {\cos\left({2\pi \over n} k\right) \over k}\leq\ln n +1.
$$
and
$$
\sum_{k=1}^{n} {\cos\left({2\pi \over n} k\right) \over k} \geq \sum_{k=1}^{n} {1 \over k} - \sum_{k=1}^{n} {2\pi \over n} \geq \ln n - 2\pi
$$
where the first inequality in the last expression follows from $\cos x \geq 1-x$ which holds if $x \geq 0$.

\qed

From proposition \ref{prop:sum1} we get the following equality for the first big summand of \eqref{sumParts}:
$$
\sum_{k=1}^{\lfloor n^{1-\epsilon}\rfloor}{\cos\left(2 \pi n^{\epsilon-1} k\right)\over k} = (1-\epsilon)\ln n + O(1).
$$

We can also obtain the following bound for the third big summand of  \eqref{sumParts}:
$$
\left|\sum_{k=\lfloor n^{1-\epsilon}\lfloor n^\epsilon\rfloor\rfloor+1}^n {\cos\left(2\pi n^{\epsilon-1}k\right)\over k}\right|<{n^{1-\epsilon}+1 \over n-n^{1-\epsilon}}=o(1).
$$

We replace the second big summand of \eqref{sumParts} with
\begin{equation}
\label{part2eq1}
\sum_{l=1}^{\lfloor n^\epsilon \rfloor-1}\sum_{t=\lfloor n^{1-\epsilon} l\rfloor+1}^{\lfloor n^{1-\epsilon}l\rfloor+\lfloor n^{1-\epsilon}\rfloor}{\cos\left(2\pi n^{\epsilon-1}t\right)\over t}.
\end{equation}
The error bacause of the replacement is
$$
\left|\sum_{l=1}^{\lfloor n^\epsilon \rfloor-1} {\cos\left(2\pi n^{\epsilon-1}\lfloor n^{1-\epsilon}(l+1)\rfloor\right)\over \lfloor n^{1-\epsilon}(l+1)\rfloor}\right|<\sum_{l=1}^{\lfloor n^\epsilon \rfloor-1}{1\over n^{1-\epsilon}l}=o(1)
$$
which follows from the fact that the inequality $\left|\lfloor x \rfloor + \lfloor y \rfloor - \lfloor x+y \rfloor\right| \le 1$ holds for all $x$ and $y$.

We rewrite \eqref{part2eq1} as
$$
\sum_{t=1}^{\lfloor n^{1-\epsilon} \rfloor}\sum_{l=1}^{\lfloor n^\epsilon \rfloor-1}{\cos\left(2\pi n^{\epsilon-1}\left(\lfloor n^{1-\epsilon} l \rfloor+t\right)\right)\over \lfloor n^{1-\epsilon} l \rfloor+t}.
$$

We get rid of the floor function in the numerator of the last expression, thus, obtaining the following sum:
\begin{equation}
\label{eq1}
\sum_{t=1}^{\lfloor n^{1-\epsilon} \rfloor}\cos\left(2\pi n^{\epsilon-1}t\right) p(t)
\end{equation}
where
$$
p(t)=\sum_{l=1}^{\lfloor n^\epsilon \rfloor-1}{1\over \lfloor n^{1-\epsilon} l \rfloor+t}.
$$

Using the fact that $\left|\cos x-\cos y\right| \leq |x-y|$ holds for all $x$ and $y$, we obtain that the cosine value because of the replacement changed at most by
$$
\left|2\pi n^{\epsilon-1}\left(\lfloor n^{1-\epsilon}l\rfloor-n^{1-\epsilon}l\right)\right|\le 2\pi n^{\epsilon-1}.
$$

Thus, we obtain the following bound of the error of the replacement:
$$
2\pi n^{\epsilon-1}\sum_{t=1}^{\lfloor n^{1-\epsilon} \rfloor}p(t)\le 2\pi n^{\epsilon-1} n^{1-\epsilon}p(t)=o(1)
$$
where we used
$$
p(t)\leq \sum_{l=1}^{\lfloor n^\epsilon \rfloor-1}{1\over n^{1-\epsilon} l}<{\epsilon \ln n+1 \over n^{1-\epsilon}}=o(1).
$$

To prove that the expression \eqref{eq1} is $O(1)$, first we will pair almost all of it's summands so that the sum of cosine values in each pair is very close to $0$.

Let $k(t)=\lfloor {n^{1-\epsilon} \over 2} \rfloor-t$ and $r(t)=\lfloor {3 n^{1-\epsilon} \over 2} \rfloor-t$. We replace \eqref{eq1} with
$$
\sum_{t=3}^{\lfloor {n^{1-\epsilon} \over 4} \rfloor-3}\left(\cos\left(2\pi n^{\epsilon-1}t\right)p(t)+\cos\left(2\pi n^{\epsilon-1}k(t)\right)p(k(t))\right)
$$
\begin{equation}
\label{part2eq2}
+\sum_{\lfloor {3n^{1-\epsilon} \over 4} \rfloor+3}^{\lfloor n^{1-\epsilon}\rfloor-3}\left(\cos\left(2\pi n^{\epsilon-1}t\right)p(t)+\cos\left(2\pi n^{\epsilon-1}r(t)\right)p(r(t))\right)
\end{equation}
where we removed some of the summands of \eqref{eq1}. Let the number of the removed summands be $C=O(1)$. From $p(t)=o(1)$ we get that the error of the last replacement is $o(1)$.

Now we replace \eqref{part2eq2} with
$$
\sum_{t=3}^{\lfloor {n^{1-\epsilon} \over 4} \rfloor-3}\left(\cos\left(2\pi n^{\epsilon-1}t\right)p(t)+\cos\left(\pi-2\pi n^{\epsilon-1}t\right)p(k(t))\right)
$$
\begin{equation}
\label{part2eq3}
+\sum_{\lfloor {3n^{1-\epsilon} \over 4} \rfloor+3}^{\lfloor n^{1-\epsilon}\rfloor-3}\left(\cos\left(2\pi n^{\epsilon-1}t\right)p(t)+\cos\left(3\pi-2\pi n^{\epsilon-1}t\right)p(r(t))\right).
\end{equation}
The error of the last replacement is $n^{1-\epsilon} \cdot 2\pi n^{\epsilon-1} \cdot o(1)=o(1)$ where the first factor is larger than the number of summands of the last sum; the second factor is the maximum change in the value of the cosine function; the third factor is $p(t)=o(1)$.

Now we can bound the maximum value of \eqref{part2eq3} with
$$
{\lfloor n^{1-\epsilon}\rfloor-C  \over 2}\sum_{l=1}^{\lfloor n^\epsilon \rfloor-1}{1\over  n^{1-\epsilon} l }-{\lfloor n^{1-\epsilon}\rfloor -C\over 2}\sum_{l=1}^{\lfloor n^\epsilon \rfloor-1}{1\over  n^{1-\epsilon} (l+1) }
$$
$$
={\lfloor n^{1-\epsilon}\rfloor-C  \over 2}\left({1\over n^{1-\epsilon}}-{1\over  n^{1-\epsilon} \lfloor n^\epsilon \rfloor }\right)=O(1).
$$
\qed

%---------------------------------------------------------------
% Section: Bibliography
%---------------------------------------------------------------

%---------------------------------------------------------------


\begin{thebibliography}{MMMMM}

\bibitem[Amb03]{Amb03}
A. Ambainis.
Quantum walks and their algorithmic applications.
{\em International Journal of Quantum Information} 1:507-518, 2003.

\bibitem[Amb04]{Amb04}
A. Ambainis.
Quantum walk algorithm for element distinctness. 
{\em SIAM J. Comput. 37(1)}, 210-239, 2007.
2001.

\bibitem[AKR05]{AKR05}
A. Ambainis, J. Kempe, A. Rivosh.
Coins make quantum walks faster,
{\em Proceedings of SODA'05}, 1099-1108, 2005.

\bibitem[AR08]{AR08}
A. Ambainis, A. Rivosh.
Quantum random walks with multiple or moving marked locations,
{\em Proceedings of SOFSEM'08}, 485-496, 2008.

\bibitem[BS06]{BS06}
H. Buhrman and R. Spalek.
Quantum Verification of Matrix Products.
{\em Proceedings of 17th Annual ACM-SIAM Symposium on Discrete Algorithms (SODA'06)}, 880-889, Miami, Florida, 2006.

\bibitem[BV]{BV}
E. Bernstein, U. Vazirani, Quantum complexity theory.
{\em SIAM Journal on Computing,} 26:1411-1473, 1997.

\bibitem[CC+03]{CC+03}
A.M. Childs, R. Cleve, E. Deotto, E. Farhi, S. Gutmann, and D. A. Spielman. Exponential algorithmic speedup by a quantum walk.
{\em Proceedings of the 35th ACM STOC}, 59-68, 2003.

\bibitem[CG04]{CG04}
A. Childs, J. Goldstone.
Spatial search and the Dirac equation, 
{\em Physical Review A}, 70:042312, 2004. 

\bibitem[Gro96]{Gro96}
L. Grover. 
A fast quantum mechanical algorithm for database search. 
{\em Proceedings of the 28th ACM STOC}, 212-219, Philadelphia, Pennsylvania, 1996. ACM Press.

\bibitem[Kem03]{Kem03}
J. Kempe. 
Quantum random walks - an introductory overview. 
{\em Contemporary Physics}, 44(4):302-327, 2003.

\bibitem[KM+10]{KM+10}
H. Krovi, F. Magniez, M. Ozols, J. Roland
Finding is as easy as detecting for quantum walks.
{\em ICALP'10 Proceedings of the 37th international colloquium conference on Automata, languages and programming.}

\bibitem[Mey96]{Mey96}
David A. Meyer. 
From quantum cellular automata to quantum lattice gases. 
{\em Journal of Statistical Physics 85}, 551-574, 1996.

\bibitem[MPA10]{MPA10}
F. L. Marquezino, R. Portugal, G. Abal.
Mixing times in quantum walks on two-dimensional grids.
arxiv:1006.4625.

\bibitem[MSS05]{MSS05}
F. Magniez, M. Santha, M. Szegedy.
An $O(n^{1.3})$ quantum algorithm for the triangle problem.
{\em Proceedings of SODA'05 1109-1117(2005), SIAM J. Comput}. 37(2): 413-424, 2007.

\bibitem[SKW03]{SKW03}
N. Shenvi, J. Kempe, and K.B. Whaley.
A quantum random walk search algorithm. 
{\em Physical Review A}, 67(5):052307, 2003.

\bibitem[Sze04]{Sze04}
M. Szegedy.
Quantum speed-up of Markov Chain based algorithms.
{\em Proceedings of IEEE FOCS'04}, 32-41.

\bibitem[Tul08]{Tul08}
A. Tulsi.
Faster quantum-walk algorithm for the two-dimensional spatial search.
{\em Phys. Rev. A} 78, 012310 (2008)

\end{thebibliography}
\end{document}